\documentstyle{article}\begin{document}\centerline{\bf Solvable Sextic
Equations}\vskip .3in \centerline{ C. Boswell$^{1}$ and M.L.
Glasser$^{1,2}$}\vskip .2in \centerline{$^1$Department of
Mathematics and Computer Science} \centerline{Clarkson University}
\centerline{Potsdam, NY 13699-5820} \vskip .1in \centerline{$ ^2$
Departamento de F\'sica Te\'orica, At\'omica y \'Optica}
\centerline{Facultad de Ciencia, Universidad de Valladolid}
\centerline{47005 Valladolid, Spain} \vskip .5in
\begin{quote}
Criteria are given for determining whether an irreducible sextic
equation with rational coefficients is algebraically solvable over
the complex number field $\cal{C}$.
\end{quote}
\vskip .5in\noindent CN: 12D10, 12E12

\vskip .5in\noindent Keywords: Sextic equation, Galois group,
resolvent equation.
\newpage
\centerline{\bf Inroduction}\vskip .1in

All equations of degree 4 or less have algebraic solutions over
${\cal{C}}$, but from the work of Ruffini, Abel and Galois it is
now well-known that all quintic equations do not[1]. One of the
first to examine this more closely was D.S. Dummit [2], who in
1991 provided a sixth degree resolvent equation for any quintic
and proved that a quintic equation  with rational coefficients is
solvable if and only if its resolvent has a rational root. He then
went on to classify the solvable quintics by giving a procedure
for determining their Galois groups. Three years later B.K.
Spearman and K.S. Williams [3] used Dummit's results to provide
the specific algebraic solution to any solvable quintic. Their
proof provides a generalization of Cardano's method for the cubic
equation. Specifically, in terms of the Bring-Jerrard form for the
general quintic:
\begin{quote} The irreducible polynomial
$$x^5+ax+b=0$$
with rational coefficients is solvable by radicals if and only if
there exist rational numbers $\epsilon=\pm1$, $c>0$ and $e\ne0$
such that
$$a=\frac{5e^4(3-4\epsilon c)}{c^2+1}\mbox{\hskip .2in and \hskip
.2in}b=\frac{-4e^5(11\epsilon+2c)}{c^2+1}$$ in which case the
roots are
$$x_j=e\sum_{k=1}^4\omega^{jk}u_k,\mbox{\hskip .3in} j=0,1,2,3,4$$
where $\omega=exp(2\pi i/5)$ and
$$u_1=(v_1^2v_3/D^2)^{1/5},\mbox{\hskip .2in} u_2=(v_3^2v_4/D^2)^{1/5},$$
$$ u_3=(v_2^2v_1/D^2)^{1/5},\mbox{\hskip .2in}u_4=(v_4^2v_2/D^2)^{1/5}. $$
Here, $D=c^2+1$, $v_1=\sqrt{D}+\sqrt{D-\epsilon\sqrt{D}}$,
$v_2=-\sqrt{D}-\sqrt{D+\epsilon\sqrt{D}}$,
$v_3=-\sqrt{D}+\sqrt{D+\epsilon\sqrt{D}}$ and
$v_4=\sqrt{D}-\sqrt{D-\epsilon\sqrt{D}}.$.
\end{quote}

Solvable irreducible quintic equations are  rare. For example,
with $-40\le a,b\le 40$, apart from the cases where $a=0$ and $b$
is not a fifth root, there are only six: $x^5+20x\pm 32=0$,
$x^5+15x\pm12=0$ and $x^5-5x\pm12=0$. The aim of this work is to
extend Dummit's analysis to equations of degree six; it is
reasonable to expect that solvable irreducible sextics are also
rare.

\vskip .2in \centerline{\bf Galois Approach to Degree Six}\vskip
.1in

As is well known, by means of a Tchirnhausen transformation the
general sextic
$$ x^6+ax^5+bx^4+cx^3+dx^2+ex+f=0\eqno(1)$$
can, in principle, be reduced to the form
$$p(x)=x^6+x^2+dx+e=0\eqno(2)$$
having roots $u_j$, $j=1,2,3,4,5,6$. Furthermore, any irreducible
equation (2) with rational coefficients is solvable by radicals if
and only if its Galois group $G$ is a subgroup of one of the two
transitive subgroups of $S_6$: $J=S_2\#S_3$ and $K=S_3\#Z_2$. Here
we use $\#$ to denote the wreath product of two groups [4]. \vskip
.1in \noindent {\bf Case 1. $G\le J$.} \vskip .1in

The group $J$ has order 48 and index 15 in $S_6$ and is generated
by the three permutations: (123)(456), (12)(45) and (14). It is
straightforward to find the symmetric function
$$\theta_1=u_1u_2u_3^2u_4u_5u_6^2+u_1u_2^2u_3u_4u_5^2u_6+u_1^2u_2u_3u_4^2u_5u_6\eqno(3)$$
which is invariant under $J$ (for which $J$ is the stabilizer).
Its conjugates in $S_6$ are
$$\mbox{in} (45)J\mbox{\hskip
.2in}\theta_2=u_1u_2u_3^2u_4u_5u_6^2+u_1u_2^2u_3u_4^2u_5u_6+u_1^2u_2u_3u_4u_5^2u_6$$
$$\mbox{in} (35)J\mbox{\hskip
.2in}\theta_3=u_1u_2u_3^2u_4u_5u_6^2+u_1^2u_2^2u_3u_4u_5u_6+u_1u_2u_3u_4^2u_5^2u_6$$
$$\mbox{in} (56)J\mbox{\hskip
.2in}\theta_4=u_1u_2u_3^2u_4u_5^2u_6+u_1^2u_2u_3u_4^2u_5u_6+u_1u_2^2u_3u_4u_5u_6^2$$
$$\mbox{in} (46)J\mbox{\hskip
.2in}\theta_5=u_1u_2u_3^2u_4^2u_5u_6+u_1^2u_2u_3u_4u_5u_6^2+u_1u_2^2u_3u_4u_5^2u_6$$
$$\mbox{in} (26)J\mbox{\hskip
.2in}\theta_6=u_1u_2^2u_3^2u_4u_5u_6+u_1^2u_2u_3u_4^2u_5u_6+u_1u_2u_3u_4u_5^2u_6^2$$
$$\mbox{in} (34)J\mbox{\hskip
.2in}\theta_7=u_1^2u_2u_3^2u_4u_5u_6+u_1u_2^2u_3u_4u_5^2u_6+u_1u_2u_3u_4^2u_5u_6^2$$
$$\mbox{in} (16)(24)J\mbox{\hskip
.2in}\theta_8=u_1^2u_2u_3^2u_4u_5u_6+u_1u_2^2u_3u_4u_5u_6^2+u_1u_2u_3u_4^2u_5^2u_6\eqno(4)$$
$$\mbox{in} (15)(34)J\mbox{\hskip
.2in}\theta_9=u_1u_2u_3^2u_4u_5^2u_6+u_1^2u_2^2u_3u_4u_5u_6+u_1u_2u_3u_4^2u_5u_6^2$$
$$\mbox{in} (13)(45)J\mbox{\hskip
.2in}\theta_{10}=u_1u_2u_3^2u_4u_5^2u_6+u_1^2u_2u_3u_4u_5u_6^2+u_1u_2^2u_3u_4^2u_5u_6$$
$$\mbox{in} (24)(35)J\mbox{\hskip
.2in}\theta_{11}=u_1u_2u_3^2u_4^2u_5u_6+u_1^2u_2^2u_3u_4u_5u_6+u_1u_2u_3u_4u_5^2u_6^2$$
$$\mbox{in} (23)(45)J\mbox{\hskip
.2in}\theta_{12}=u_1u_2u_3^2u_4^2u_5u_6+u_1^2u_2u_3u_4u_5^2u_6+u_1u_2^2u_3u_4u_5u_6^2$$
$$\mbox{in} (26)(45)J\mbox{\hskip
.2in}\theta_{13}=u_1u_2^2u_3^2u_4u_5u_6+u_1u_2u_3u_4u_5^2u_6+u_1u_2u_3u_4^2u_5u_6^2$$
$$\mbox{in} (26)(15)J\mbox{\hskip
.2in}\theta_{14}=u_1u_2^2u_3^2u_4u_5u_6+u_1^2u_2u_3u_4u_5u_6^2+u_1u_2u_3u_4^2u_5^2u_6$$
$$\mbox{in} (26)(34)J\mbox{\hskip
.2in}\theta_{15}=u_1^2u_2u_3^2u_4u_5u_6+u_1u_2^2u_3u_4^2u_5u_6+u_1u_2u_3u_4u_5^2u_6^2$$

Since $J$ fixes $\theta_1$ and permutes its conjugates among
themselves, the polynomial
$$f(x)=\prod_{i=1}^{15}(x-\theta_i)\eqno(5)$$
is J-invariant. If $G\le J$, then $\theta_1$ is invariant under
all the automorphisms of $G$ and is therefore rational. On the
other hand, if $G$ is not a subgroup of $J$, then $G$ contains an
automorphism which does not leave $\theta_1$ invariant, so
$\theta_1$ will not be rational. Thus, $G$ is $J$ or a subgroup of
$J$ if and only if $f(x)$ has at least one root fixed by $J$, i.e.
$f(x)$ has a rational root.

It remains to express the coefficients of $f(x)$ in terms of those
of $p(x)$. The former are symmetric functions of the $\theta_j$,
which are in turn symmetric functions of the roots $u_k$. Hence,
the coefficients of $f(x)$ are expressible in terms of the
elementary symmetric functions of the roots, which are precisely
the coefficients of $p(x)$.

As usual, we denote the elementary symmetric function in $y_1,
\dots, y_k$ of degree $n$ by $\sigma_n(y_1,\dots, y_k))$ and we
seek the expression of $\sigma_n(\theta_1,\dots,\theta_{15})$ in
terms of
$\sigma_1(u_1,\dots,u_6)=0=\sigma_2(u_1,\dots,u_6)=\sigma_3(u_1,\dots,u_6)$,
$\sigma_4(u_1,\dots,u_6)=1$, $\sigma_5(u_1,\dots,u_6)=-d$ and
$\sigma_6(u_1,\dots,u_6)=e$. By treating these as functional
identities, we simply chose values for the roots to obtain a set
of independent linear equations. Since each $\theta_j$ is of
degree 8, $\sigma_1$ is of degree 8, $\sigma_2$ of degree 16, etc.
Thus, the symmetric functions of the $\theta$'s were expressed as
linear combinations of terms, of the same degree, of the
$\sigma_n(u_1,\dots,u_6)$. For example,
$$\sigma_1(\theta_j)=A_1\sigma_2(u_k)\sigma_6(u_k)+A_2\sigma_3(u_k)\sigma_5(u_k)+A_3\sigma_4^2(u_k)+$$
$$A_4\sigma_2^2(u_k)\sigma_4(u_k)+A_5\sigma_2(u_k)\sigma_3^2(u_k)+A_6\sigma_2^4(u_k).\eqno(5)$$
This led to sets of linear equations involving as many as 149
unknown $A$'s which we were able to handle using MAPLE. The
problem of expressing $\sigma_n(\theta_j)$ in terms of the roots
was treated recursively using Newton's formula. This procedure was
first carried out for the general sextic (1) which led to an
expression for $f(x)$ four pages long, which we shall not
reproduce here. For the reduced equation (2) we have

\begin{quote}
{\bf THEOREM 1}

The sextic equation $x^6+x^2+dx+e=0$ is solvable by radicals and
its Galois group is a subgroup of $J$ if and only if the resolvent
equation
$$x^{15}-6e^2x^{13}-(42e+3)e^3x^{12}+7e^4x^{11}+(222e-21d^2)e^5x^{10}+$$
$$(453e^2+57e+8)e^6x^9-(340e-109d^2)e^7x^8-(1716e^2-288d^2e+17)x^7-$$
$$(1232e^3-300e+144d^2)e^9x^6+(1534e^2+538d^2e-353d^4+2)e^{10}x^5+$$
$$(2592e^3-96d^2e^2-258e+48d^2)e^{11}x^4-(1728e^4+\eqno(6)$$
$$1012e^2-284d^2e+94d^4-9)e^{12}x^3+(432e^3-2160d^2e^2+792d^4e+118e+$$
$$5d^2)e^{13}x^2+(1296d^2e^3-27e^2+138d^2e-60d^4-4)e^{14}x+$$
$$(144d^4e-32d^6-3d^2)e^{15}=0.$$
\end{quote}

The simplest examples can be obtained simply by requiring the
constant term on the left hand side of (6) to vanish, which gives
$$e=\frac{32d^4+3}{144d^2}\eqno(7)$$
Thus, if $d=1/2$, we find that the Galois group of the irreducible
polynomial $36x^6+36x^2+18x+5$ is a subgroup of $J$ and is
solvable.

\vskip .1in\noindent {\bf Case 2. $G\le K$} \vskip .1in

It is easily verified that $K$ is the stabilizer of the symmetric
function
$$\phi_1=u_1u_2u_3^2+u_1^2u_2u_3+u_1u_2^2u_3+u_4u_5u_6^2+u_4^2u_5u_6+u_4u_5^2u_6\eqno(8)$$
of the roots. The conjugates of $\phi_1$ are
$$\phi_2=u_4u_2u_3^2+u_4^2u_2u_3+u_4u_2^2u_3+u_1u_5u_6^2+u_1^2u_5u_6+u_1u_5^2u_6\mbox{\hskip
.1 in}(14)K$$
$$\phi_3=u_5u_2u_3^2+u_5^2u_2u_3+u_5u_2^2u_3+u_4u_1u_6^2+u_4^2u_1u_6+u_4u_1^2u_6\mbox{\hskip
.1 in}(15)K$$
$$\phi_4=u_6u_2u_3^2+u_6^2u_2u_3+u_6u_2^2u_3+u_4u_5u_1^2+u_4^2u_5u_1+u_4u_5^2u_1\mbox{\hskip
.1 in}(16)K$$
$$\phi_5=u_1u_4u_3^2+u_1^2u_4u_3+u_1u_4^2u_3+u_2u_5u_6^2+u_2^2u_5u_6+u_2u_5^2u_6\mbox{\hskip
.1 in}(24)K\eqno(9)$$
$$\phi_6=u_1u_5u_3^2+u_1^2u_5u_3+u_1u_5^2u_3+u_4u_2u_6^2+u_4^2u_2u_6+u_4u_2^2u_6\mbox{\hskip
.1 in}(25)K$$
$$\phi_7=u_1u_6u_3^2+u_1^2u_6u_3+u_1u_6^2u_3+u_4u_5u_2^2+u_4^2u_5u_2+u_4u_5^2u_2\mbox{\hskip
.1 in}(26)K$$
$$\phi_8=u_1u_2u_4^2+u_1^2u_2u_4+u_1u_2^2u_4+u_3u_5u_6^2+u_3^2u_5u_6+u_3u_5^2u_6\mbox{\hskip
.1 in}(34)K$$
$$\phi_9=u_1u_2u_5^2+u_1^2u_2u_5+u_1u_2^2u_5+u_4u_3u_6^2+u_4^2u_3u_6+u_4u_3^2u_6\mbox{\hskip
.1 in}(35)K$$
$$\phi_{10}=u_1u_2u_6^2+u_1^2u_2u_6+u_1u_2^2u_6+u_4u_5u_3^2+u_4^2u_5u_3+u_4u_5^2u_3\mbox{\hskip
.1 in}(36)K$$ As above, we define the polynomial
$$g(x)=\prod_{i=1}^{10}(x-\phi_i).\eqno(10)$$
To get the coefficients of $g(x)$ we have followed the procedure
described above to express the $\sigma_n(\phi_j)$ in terms of the
$\sigma_n(u_k)$. Since the degree of $\sigma_1(u_k)$ is only 4 in
this case, the calculations are not as extensive. In this way we
have \vskip .2in
\begin{quote}
{\bf THEOREM 2}\vskip .1in

The Galois group of the reduced sextic  (2), having rational
coefficients, is a subgroup of K if and only if the tenth degree
resolvent equation
$$x^{10}+4x^9+6x^8-(66e^2-4)x^7-(324e^2-58d^2e-1)x^6-(642e^2-192d^2e+11d^4)x^5+$$
$$(129e^4-640e^2+246d^2e-22d^4)x^4+(384e^4-74d^2e^3-320e^2+144d^2e-16d^4)x^3+$$
$$(384e^4-108d^2e^3+4d^4e^2-64e^2+32d^2e-4d^4)x^2-(64e^6-128e^4-32d^2e^3+\eqno(10)$$
$$40d^4e^2-6d^6e)x-(64e^6-16d^2e^5-64d^2e^3+48d^4e^2-12d^6e+d^8)=0.$$
has a rational root.
\end{quote}
Two simple examples are furnished by $d=e=4$ and $d=2=2e$ which
give the irreducible solvable equations $x^6-x^2+4x+4=0$ and
$x^6-x^2+2x+1=0$ respectively. \vskip .1in\centerline{\bf
Discussion}

We can refine the specification of the Galois groups by looking at
the discriminant
$$\Delta=\prod_{i\ne j}(u_i-u_j).\eqno(11)$$
For the reduced equation (2) this is
$$\Delta=46656e^5+13824e^3-43200d^2e^2+22500d^4e+1024e-3125d^6-256d^2.\eqno(12)$$
Thus, $G$ is a subgroup of the alternating group $A_6$ if and only
if $\sqrt{\Delta}$ is rational. Hence if $\sqrt{\Delta}$ is
rational the Galois group of $p(x)$ is a subgroup of $L=J\cap A_6$
if and only if $f(x)$ has a rational root and it is a subgroup of
$M=K\cap A_6$ if and only if $g(x)$ has a rational root. $L$ has
index 30 in $S_6$ and is generated by $(123)(456)$, $(12)(45)$ and
$(14)(25)$. $M$ has index 20 and is generated by (123) and
(14)(25)(36). We note also that $G\le J\cap K$, which is
isomorphic to the dihedral group $D_6$ if both $f(x)$ and $g(x)$
have a rational roots.

We have worked out the resolvent equations $f(x)$ and $g(x)$ for
the general sextic (1) (with $a=0$, since this is trivially
achieved by a linear substitution) which are available on request.
By using different methods, G. W. Smith [5] has identified the
Galois groups for several families of sextic equations such as
$x^6+(t-6)x^4+(2t-2)x^3+(t+9)x^2+6x+1=0$. We have confirmed his
results that $G\le K$ in this and a few other cases by using the
extended form of $g(x)$.

\newpage
\centerline{\bf References}\vskip .1in \noindent [1] R. Bruce
King, {\it Beyond the Quartic Equation} [Birkhauser, Boston 1996]

\noindent [2] D.S. Dummit, Math. Comp {\bf{195}}, 387-401 (1991).

\noindent [3] B.K. Spearman and K.S. Williams, Amer. Math. Monthly
{\bf{101}}, 986-992 (1994).

\noindent [4] J.D. Dixon and B. Mortimer, {\it Permutation
Groups},[Springer, New York 1991].

\noindent [5] G.W. Smith, Math. Comp.{\bf{69}}, 775-796 (2000).

\end{document}